\documentclass[%
 reprint,
superscriptaddress,
 amsmath,amssymb,
 aps,
]{revtex4-2}

\usepackage{upgreek}
\usepackage{graphicx}
\usepackage{dcolumn}
\usepackage{bm}
\usepackage{xcolor} 
\usepackage[colorlinks=true, allcolors=blue]{hyperref} 

\begin{document}

\preprint{APS/123-QED}

\title{Non-reciprocal Coulomb drag in a ballistic quantum wire}

\author{Suyang Cai}
\affiliation{%
 Department of Physics, University of Florida, Gainesville, FL 32611, USA
}%
 \author{Mingyang Zheng}
 \affiliation{%
 Department of Physics, University of Florida, Gainesville, FL 32611, USA
}%
\author{Nathan Rao}
\affiliation{%
 Department of Physics, University of Florida, Gainesville, FL 32611, USA
}%
\author{Glen Gillia}
\affiliation{%
 Department of Physics, University of Florida, Gainesville, FL 32611, USA
}%
\author{Rebika Makaju}
\affiliation{%
 Department of Physics, University of Florida, Gainesville, FL 32611, USA
}%
\author{S. J. Addamane}
\affiliation{
 Center for Integrated Nanotechnologies, Sandia National Laboratories, Albuquerque, New Mexico 87185, USA
}%
\author{Dominique Laroche}%
 \email{dlaroc10@ufl.edu}
\affiliation{%
 Department of Physics, University of Florida, Gainesville, FL 32611, USA
}%



\date{\today}

\begin{abstract}

1D-Coulomb drag serves as a platform for probing electron-electron interactions in 1D systems. Under the charge fluctuation formalism, the non-reciprocal component of Coulomb drag signal in mesoscopic devices is predicted to rely on the breaking of translational invariance due to intrinsic disorder. In this work, we report the measurement of a Coulomb drag device with a ballistic drag wire, allowing us to study the drag signal in nearly pristine quantum wires. Surprisingly, a non-reciprocal component with strength comparable to that of the reciprocal component is still detected across the measured regime, despite the drag wire being ballistic. We suggest that the non-reciprocal signal arises from low energy disorder in the device, which is consistent with the evolution of the drag wire's conductance at low biases voltages and temperatures. Additionally, the non-reciprocal component of the drag signal shows a power-law temperature dependence that coincides with a diffusive model, while the reciprocal component's temperature dependence cannot be explained under the existing framework. The bias voltage dependence of the drag wire conductance is fitted into three models to extract the interaction parameter and disorder level within the wire, but none of the models provides a fully self-consistent explanation for the data.
\end{abstract}

\maketitle


\section{\label{sec:level1}Introduction}

Understanding interactions between electrons in the solid state has always been a key interest of condensed matter physics. By describing interacting electrons as fermionic quasiparticles, the Fermi liquid theory has successfully predicted numerous properties of three-dimensional and two-dimensional metals. However, quasiparticle description has been shown to break down in one-dimensional (1D) systems \cite{voit1995one}, where electron correlation becomes strong enough to make the excitations better described as collective modes rather than as the motion of 'dressed' particles. These low-energy bosonic modes are well described by the Tomonaga-Luttinger liquid (TLL) framework \cite{luttinger1963exactly,tomonaga1950remarks} and has successfully predicted exotic experimental phenomena such as spin-charge separation \cite{auslaender2005spin,jompol2009probing}, charge fractionalization \cite{Steinberg2008charge}, power-law conductance \cite{bockrath1999luttinger} and edge excitation in fractional quantum hall states \cite{chang1996observation}. As harnessing 1D transport in novel devices has gathered significant interest in recent years \cite{xiang2020one,microsoft2025interferometric,franklin2022carbon, klaassen2025realization}, understanding and quantifying electron interactions in TLL systems has become increasingly important.

\begin{figure*}[t]
\includegraphics[width=\textwidth, keepaspectratio]{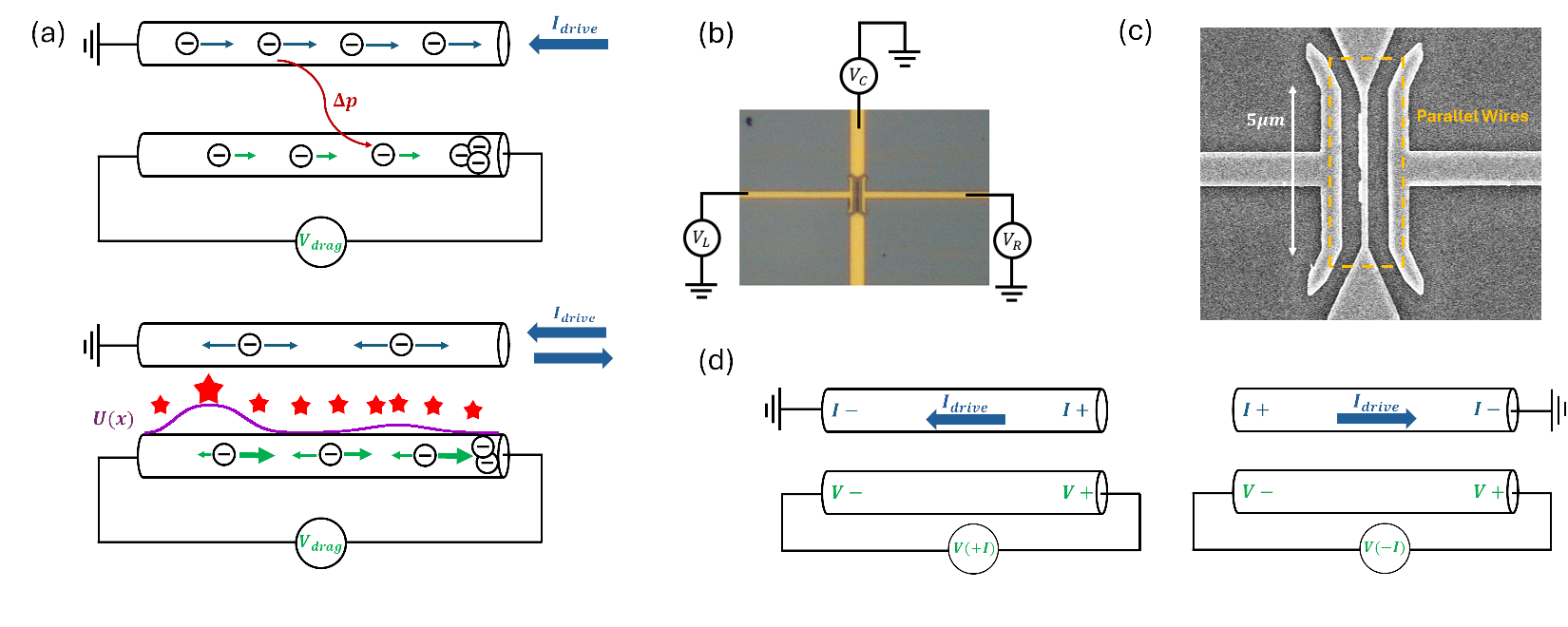}
\caption{\label{fig:epsart} (a) Schematic illustrations of the reciprocal (top) and non-reciprocal (bottom) Coulomb drag signals. For reciprocal drag, momentum transfer via Coulomb interaction results in electron accumulation along the opposite direction of the current flow, generating an anti-symmetric signal that reverses upon current direction reversal. Non-reciprocal drag is induced by asymmetric scattering due to unequal potential barriers breaking translational symmetry.  In this case, both energy fluctuations and momentum transfer from the electron-carrying current can generate a symmetrical drag signal upon current direction reversal. (b) Optical microscope figure of a typical laterally coupled lateral wire device. Negative voltages are applied to the gates to pinch off the 2DEG directly underneath the gates, confining electrons to the parallel quasi-1D channels between the gates. (c) SEM image of the red-boxed area in (b). (d) Configurations for measuring V(+I) and V(-I). The only difference in the two setups is the direction of $I_{drive}$ and the position of the ground. 
}
\label{fig1}
\end{figure*}

Among the experimental techniques used to probe 1D systems, Coulomb drag \cite{narozhny2016coulomb}, a multi-wire technique, is especially promising owing to its larger sensitivity to interaction effects than typical single-wire experiments. In a 1D Coulomb drag measurement, a current ($I_{drive}$) sourced through a quasi-1D wire induces a voltage ($V_{drag}$) across an adjacent wire in the absence of current flow, assuming negligible interwire tunneling. The traditional theoretical approach to Coulomb drag considers momentum transfer via scattering between the carriers from different wires (top panel, \autoref{fig1}(a)) as the source of the potential build up in the floating circuit, in a friction-like mechanism \cite{gramila1991mutual,klesse2000coulomb,pustilnik2003coulomb, yamamoto2006negative, laroche2011positive, laroche20141d}. The contribution from momentum transfer can be calculated by means of renormalization group method \cite{klesse2000coulomb,fuchs2005coulomb}, with back-scattering and forward-scattering both contributing to distinct temperature dependence. Within this framework, the drag signal due to momentum transfer is predicted to be reciprocal (i.e. $V_{drag}$ reverses upon $I_{drive}$ reversal, in accordance with Onsager reciprocity \cite{onsager1931reciprocal}). 

Despite the fact that momentum transfer model’s prediction aligns well with some experimental observations such as the non-monotonic temperature dependence of $V_{drag}$ \cite{laroche20141d,makaju2024nonreciprocal}, this framework does not readily account for disorder effects occurring in realistic devices. For instance, the violation of Onsager relations previously reported \cite{makaju2024nonreciprocal, zheng2025tunable, tabatabaei2020andreev, tang2020frictional} cannot be properly understood within the frictional drag formalism. A charge fluctuation model, on the other hand, offers an alternative physical approach to Coulomb drag which can explain these observations \cite{nazarov1998current,kamenev1995coulomb,levchenko2008coulomb,sanchez2010mesoscopic,kaasbjerg2016correlated, borin2019coulomb}. In this framework, a time and space dependent scalar potential originating from the charge and thermal fluctuation in the drive wire is considered, which will result in a second or third order rectification effect in the drag wire (bottom panel, \autoref{fig1}(a)). Noticeably, a nonlinear dispersion relation near the Fermi surface is essential for the generation of a nonzero signal; otherwise, the contributions from electrons and holes will cancel out each other exactly \cite{kamenev1995coulomb}. As the model accounts for all the random electric field in the system, broken translational invariance in disordered mesoscopic systems can lead to asymmetric scattering of electrons, and thus contribute to a nonreciprocal drag signal.  

Recently, numerous experiments have reported on this nonreciprocal drag in a variety of platform, including 1D-2D platforms, laterally coupled quantum wires and vertically coupled quantum wires \cite{tang2019long, anderson2021coulomb,zheng2025quasi,zheng2025tunable,makaju2024nonreciprocal}. Notably, the one-dimensional systems reported in these experiments all exhibited non-ballistic conductance owing to the presence of disorder, providing a mechanism for spatial symmetry breaking and the generation of a nonreciprocal drag signal. However, in clean ballistic quantum wires where spatial symmetry is conserved, one would expect only reciprocal Coulomb drag signals to be generated. In this paper, we present the Coulomb drag signal measured in a 7.2-micron long quasi-1D wire with ballistic conductance, allowing us to investigate Coulomb drag in a regime with minimal impurity scattering. Even in the ballistic regime, both reciprocal and non-reciprocal drag signals of comparable magnitude are reported. These results suggest a model for ballistic transport in 1D system where unavoidable symmetry breaking in the quantum wires occurs, either through low-energy scale disorder or through unavoidable asymmetric energy barriers generated at the interface between the 2D and the 1D section of the device \cite{lal2001transport}.



\begin{figure*}[t]
\includegraphics[width=\textwidth]{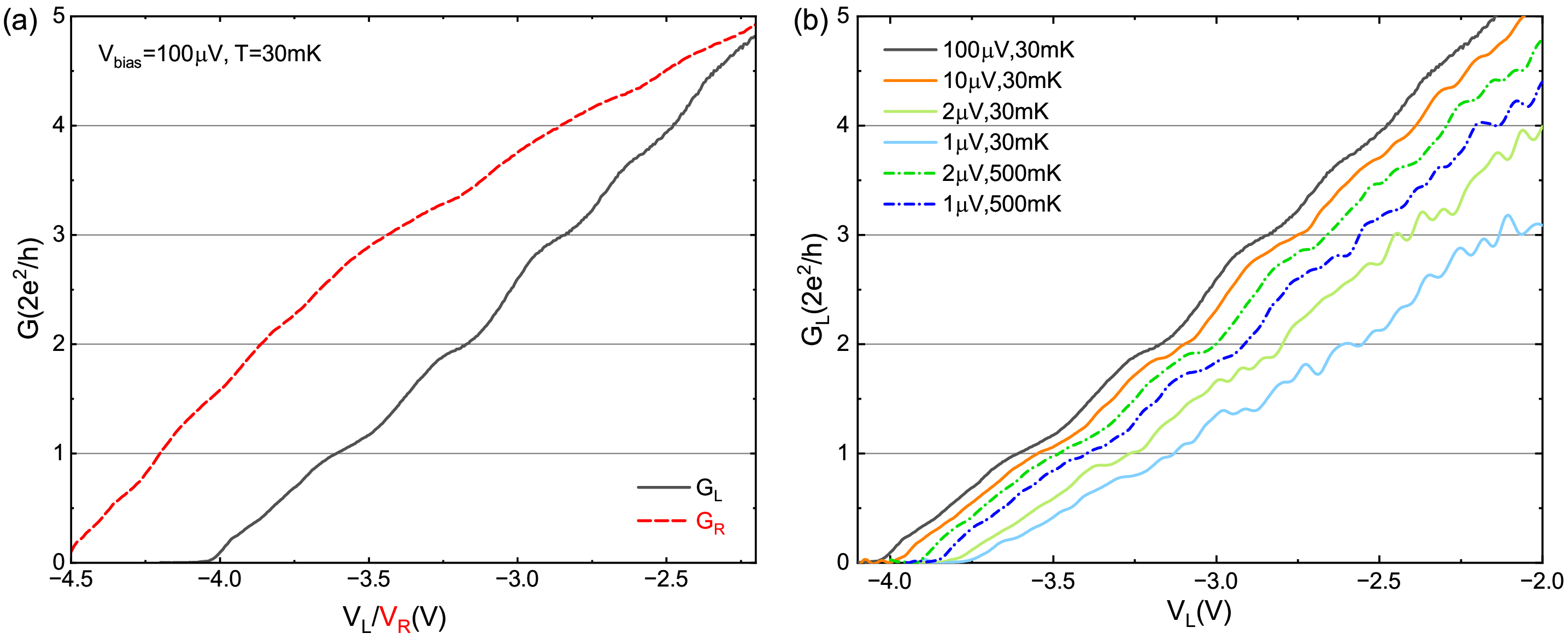}
\caption{\label{fig:wide} Quantum wires conductance. (a)  Conductance of the left (right) wire as a function of the left (right) gate voltage at base temperature ($T=30 \text{ mK}$) and under $V_{bias}=100 \text{ $\upmu$V}$, where left wire conductance is shown in solid black curve and right wire is shown in dashed red curve. The left wire is ballistic, with plateaus appearing at integer multiples of ($\frac{2e^2}{h}$), while the right wire is non-ballistic.
(b) Conductance of the left wire as a function of the left gate voltage under various temperatures and bias voltages conditions. Conductance deviates from ballistic transmission at low temperature and low bias voltage. At 30 mK, the 2 $\upmu$V and 1 $\upmu$V curves have been smoothed, as the large fluctuations hindered the identification of the average conductance within a plateau. Linetraces have been offsetted in gate voltage by 0.05V for visibility purposes.
}
\label{fig2}
\end{figure*}

\section{Device Fabrication and operation}
Our device is fabricated on a GaAs/AlGaAs heterostructure hosting a two-dimensional electron gas (2DEG). The active region is defined by a mesa electrically contacted by Ge/Au/Ni/Au ohmic contacts. Three Ti/Au top gates are patterned by e-beam lithography and deposited using standard metal-evaporation techniques. The top-down optical and scanning electron microscope (SEM) picture of a typical parallel-wires arrangement are shown in \autoref{fig1}(b) and (c), respectively. The center gate is negatively biased so that interwire tunnelling is suppressed ($R_{tunnel} > 25 \text{ M}\Omega$) and two independently contacted wires can be created, enabling Coulomb drag measurements. The left and right gates are applied with various negative voltage to control independently the respective wire's width, effectively tuning the Fermi energy in each wire (and thus the number of quasi-1D channels). All the measurements presented in this article are performed in a Bluefors dilution refrigerator using standard lock-in technique with a frequency of 37 Hz. The Coulomb drag measurements were performed using a constant AC current of 10 nA, unless otherwise stated. More detailed fabrication procedures and characterization details can be found in supplementary sections 2 and 3, and in Ref. \cite{makaju2024nonreciprocal}.

\begin{figure*}[t]
\includegraphics{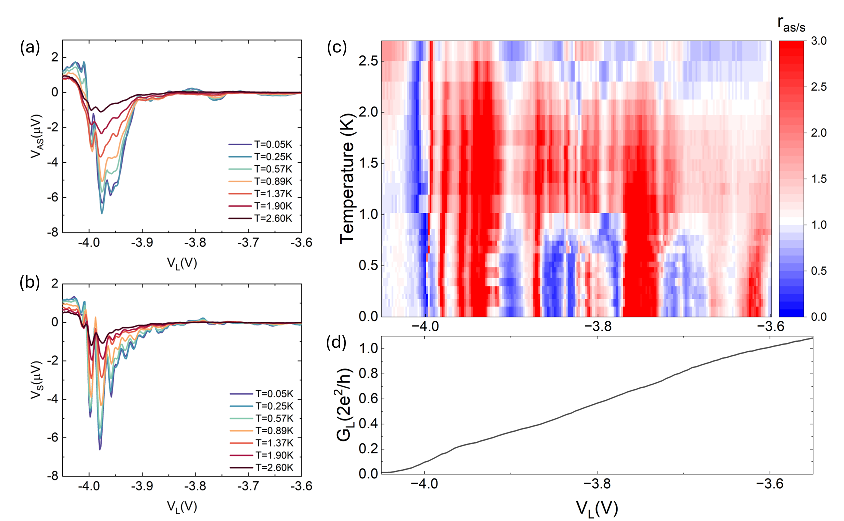}
\caption{\label{fig:wide} Coulomb drag signal.  $V_L$ dependence of (a) the anti-symmetric ($V_{AS}$) and (b) the symmetric ($V_{S}$) drag voltage at various temperatures. The signal generally decreases as the temperature increases, and the two components of the signal have comparable peak amplitudes. The symmetric component has more oscillations due to an apparent higher sensitivity to disorder. (c) $V_L$ and temperature dependence of $r_{AS/S}$, the ratio between reciprocal and non-reciprocal signal. The anti-symmetric component dominates over a wider regime at higher temperatures, as previously reported \cite{zheng2025tunable}. (d) $V_L$ dependence of $G_L$. The antisymmetric signal dominates whenever a plateau or shoulder in the conductance $G_L$ is observed. These features are attributed to disorder in the low-density wire within the single-subband regime.
}
\label{fig3}
\end{figure*}

\begin{figure*}[t]
\includegraphics[width=\textwidth]{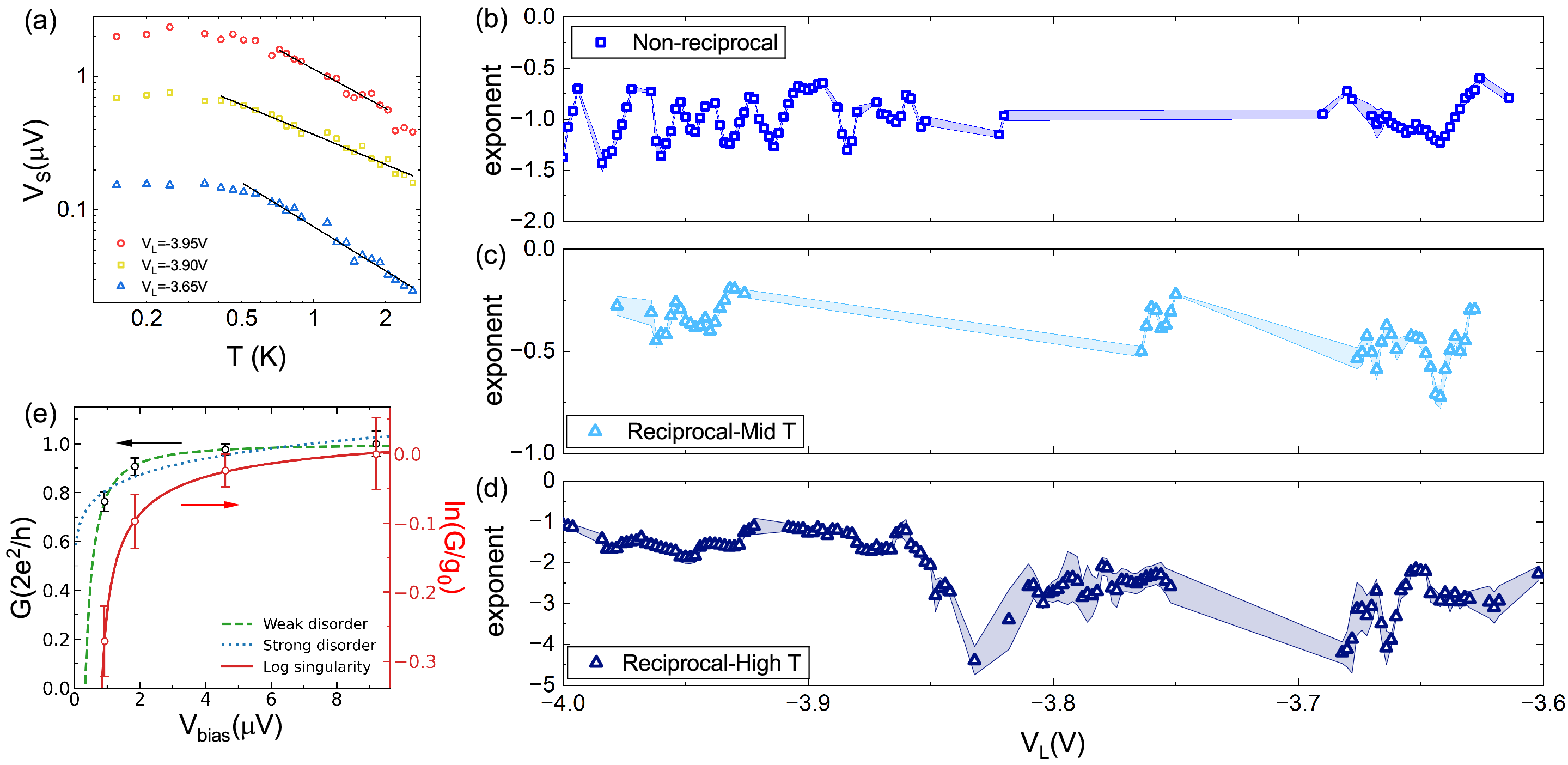}
\caption{\label{fig:wide} Fitting for temperature dependence of signal components. Bad fittings with $R^2<0.85$ or multiple points comparable with noise level are omitted and can be seen supplementary section IV. (a) Representative power-law fittings for the non-reciprocal signal for $V_L--3.95\text{ V}$ (red circle), $V_L--3.90\text{ V}$ (yellow square) and $V_L--3.65\text{ V}$ (blue triangle). (b) $V_L$ dependence of power-law fitting parameters for non-reciprocal signal, where the error bars are the maximum change in slope if the final fitting range is narrowed by one data point from either the left or right end (the same as the error bars in (c) and (d)). The fitting parameters are clustered around -1. (c) $V_L$ dependence of power-law fitting parameters for the reciprocal signal in the mid-temperature regime (300 mK $\sim$ 800 mK). The fitting parameters are clustered around -0.33 at larger negative voltages and around -0.49 at smaller negative voltages. (d) $V_L$ dependence of power-law fitting parameters for the reciprocal signal in the high-temperature regime (1 K$\sim$2.6 K). The exponent shows notable dependence on the gate voltage value. (e) Left wire conductance of the first plateau vs bias voltage. Fitting curve for three different models are also shown.
}
\label{fig4}
\end{figure*}

\section{Measurement Results}

The conductance of the left and the right wires, $G_L$ and $G_R$, as a function of their respective plunger gate voltage $V_{L}$ and $V_{R}$, are shown in \autoref{fig2}(a). Notably, the left wire exhibits ballistic conductance for $N\leq3$ (where N is the number of occupied subbands), as evidenced by conductance plateaus at integer multiples of {$2e^2/h$}. In contrast, the right wire does not show ballistic conductance plateaus, with only minor shoulders visible in $G_R$. Given the minimal disorder scattering, the left wire is designated as the drag (passive) wire for Coulomb drag measurements presented in this work, while the right wire serves as the drive (active) wire. 

\autoref{fig2}(b), shows the ballistic left wire conductance for different bias voltage and temperature conditions.  $G_L$ deviates significantly from the ballistic behavior at lower bias voltage, while elevated temperature leads to the recovery of idealized plateau value. Such observations are consistent with previously reported experiments in ballistic wires \cite{tarucha1995reduction, yacoby1996nonuniversal, levy2006luttinger} and has been attributed to the presence of weak disorder \cite{maslov1995transport} or unavoidable potential barriers at the transition between the 2D and the 1D regime in physical devices \cite{lal2001transport}.

To understand the physical mechanism of the drag signal, we extracted the reciprocal (antisymmetric) and non-reciprocal (symmetric) components using the following procedure \cite{zheng2025tunable}. First, the drag signal $V(+I)$ is measured with an AC drive current $I_{drive}=10 \text{ nA}$ (\autoref{fig1}(d), left panel). Then, the current is reversed by exchanging the drive current probes while leaving the voltage probes unchanged, and the drag signal $V(-I)$ is measured (\autoref{fig1}(d), right panel). The reciprocal (antisymmetric, AS) and non-reciprocal (symmetric, S) signals are then calculated $V_{AS}(I)=[V(+I)-V(-I)]/2$ and $V_S(I)=[V(+I)+V(-I)]/2$, respectively. The antisymmetric component of the drag signal under fixed $V_{R}=-4.15 \text{ V}$ at various temperatures is shown in \autoref{fig3}(a), while the corresponding symmetric component is shown in \autoref{fig3}(b). It is worth noticing that the $V_{L}$ range shown in \autoref{fig3}(a) and \autoref{fig3}(b) is within the single subband regime (i.e. from the pitch-off voltage to the fully occupation of the first subband). As electron density exceeds one-subband regime, the signal strength becomes comparable with the noise level owing to the large ($\sim 200$ nm) interwire separation.  As such, this work focuses on the single-subband regime. Both the reciprocal and the non-reciprocal contributions show oscillations of comparable amplitudes, with several sign changes across the gate voltage range. As anticipated, $V_S$ oscillates more rapidly than $V_{AS}$. This is consistent with the rectification origin, which suggests that the non-reciprocal signal is dependent on the detailed potential landscape across the drag wire. While early theoretical work predicted a drag signal monotonically increasing with temperature under the charge-fluctuation model \cite{levchenko2008coulomb}, our experiment show a decreasing drag signal (sometimes non-monotonically) with increasing temperature for both $V_S$ and $V_{AS}$, in agreement with past experimental observations \cite{makaju2024nonreciprocal, zheng2025tunable} and highlighting the presence of strong electron-electron interactions in the coupled 1D system.

To quantify the relative strength of $V_{AS}$ and $V_S$, we calculate the ratio $r_{AS/S}=\frac{|V_{AS}|+\Gamma}{|V_S|+\Gamma}$. Here, $\Gamma=0.1$ $\upmu$V is comparable to the noise level and is applied to ensure $r_{AS/S}$ approaches a value of 1 when both $V_{AS}$ and $V_S$ have small amplitudes. The gate voltage and temperature dependence of $r_{AS/S}$ is shown in \autoref{fig3}(c). Compared to a similar device with non-ballistic conductance \cite{zheng2025tunable}, the ballistic device exhibits a significantly wider gate voltage regime where reciprocal signal dominates over non-reciprocal signal. This domination widens at higher temperature ($T>1$ K) until T $\sim$ 2.4 K, where the signal is comparable to the noise level. This observation aligns well with the expectation that disorder-induced scattering is a primary mechanism inducing the non-reciprocal signal as disorder effects get smeared out at elevated temperatures. It is also worth noting that the gate voltage range where the reciprocal component dominates coincides with the appearance of ‘shoulders’ in left wire conductance at densities well below the first subband value (shown in \autoref{fig3}(d)). 

To clarify the microscopic origin of the drag signal, we extracted the power law fitting parameters of the temperature dependence of both the non-reciprocal (\autoref{fig4}(b), representative fitting in \autoref{fig4}(a)) and reciprocal signals (\autoref{fig4}(c)(d)).  For the non-reciprocal signal, fitted in the range $0.4\text{ K}<T<2.6\text{ K}$, the extracted exponents are clustered around -1 across the entire measured $V_R$ range, indicating that $V_{drag}\propto T^{-1}$. Surprisingly, this temperature dependence is very similar with the results reported for a non-ballistic device \cite{makaju2024nonreciprocal}, where the $T^{-1}$ dependence was predicted to arise from of a three-particle correlation model in the multi-subband limit of diffusive wires, in stark contrast with the single subband ballistic quantum wire under study in this letter. On the other hand, the reciprocal signal exhibits two distinct temperature dependence regimes, with different power-law exponents than those observed in the nonreciprocal regime. In mid-temperature regime ($300\text{ mK}<T<800\text{ mK}$), the temperature dependence has an exponent of $\sim -0.33$ under larger negative gate voltage and an exponent of $\sim -0.49$ under lower negative gate voltage, while the exponent in the high temperature regime varies between -1 and -4 non-monotonically. The momentum-transfer model predits $V_{drag}\propto T^{2K_c -1}$ for 1D-Coulomb drag with two matched density wires, where $0<K_c<1$ is the relative Luttinger liquid (LL) interaction parameter. From the exponent extracted from the mid-T regime, $K_c$ between 0.25 and 0.33 are calculated, consistently with repulsive interactions within the Luttinger liquid framework as $K_c<1$. This value is in qualitative agreement with previously reported value for closely separated wire in GaAs \cite{tarucha1995reduction}. However, the exponent extracted in the high temperature regime cannot be readily explained by existing theories since the fit results yield unphysical negative $K_c$ values, similarly to what was observed at high temperature in vertically-coupled disordered quantum wires \cite{zheng2025tunable, zheng2026quasi}. We note that only points yielding conclusive power-law dependence are included in \autoref{fig4}(b)(c)(d). Additional information regarding the point selection and the fitting procedure is available in the supplementary Section 4. We also point out that, owing to the limited number of points in the high-temperature regime, we cannot rule out that the drag signal follows an Arrhenius temperature dependence. The result of such an analysis is presented in Supplementary section 4, but we point out that no known theoretical models predict an Arrhenius-like increase of the drag signal with decreasing temperature. 

\section{Discussion}

Despite the reciprocal signal predominance over a wide parameter ranges, the existence of a non-reciprocal contribution in a ballistic system is still worth investigating, as the breaking of translational invariance by disorder scattering is typically considered essential for the formation of non-zero rectification contribution in a mesoscopic system.  We propose that this discrepancy arises from the low-energy potential barriers within the drag wire, which scatter low-energy electrons while remaining nearly transparent to high-energy carriers. The behavior of left wire conductance under different bias voltage and temperature (\autoref{fig2}(b)) is consistent with this assumption: For biases below $2 \text{ $\upmu$V}$, the conduction is significantly lower than the ideal quantized value ($2Ne^2/h$), decreasing further as the bias is lowered. However, upon raising the temperature from base temperature (30 mK) to 500 mK, the conductance recovers the ballistic value. This behavior indicates that low-energy electrons are susceptible to scattering in the system, while the electrons with higher energy, either due to higher bias voltage or thermal excitation under higher temperature, have near ballistic transmission through the wire. It is especially relevant within the charge fluctuation model, as low-energy fluctuations are statistically more likely to occur, and such fluctuations would represent the main source of nonreciprocal drag in the system. This interpretation is also in agreement with the fact that the reciprocal drag signal dominates over most of the measured voltage in high temperature regime.

To better characterize the disorder in our wire, the bias-dependent conductance data presented in \autoref{fig2}(b) is compared with theoretical predictions for conductance through disordered wires. Assuming Fermi-liquid leads, the overall conductance magnitude generally depends on the wire length, while the temperature dependence typically reflects the interaction parameters of the Luttinger liquid within a single wire.

To systematically evaluate the origin of the observed suppression, we fit the data to three candidate scaling forms: 
(i) a power-law interaction model appropriate for weak disorder, 
(ii) a power-law dependence arising from barrier-like strong disorder, and 
(iii) a logarithmic singularity model near the strong localization limit. 
All fits were performed over the same bias range using nonlinear least-squares regression, considering spinful electrons. For the logarithmic singularity model,

\begin{equation}
\ln\!\left(\frac{G}{g_0}\right) 
= A - B (V - V_c)^{-1/2},
\end{equation}

where $G$ is the conductance, $g_0 = 2e^2/h$, and $A$ and $B$ are fitting constants. 
The parameter $V_c$ defines a characteristic localization scale,

\begin{equation}
T_c = \frac{e V_c}{k_B}.
\end{equation}

For the first subband, this fit yields $T_c = 1.34\,\mathrm{mK}$. The interaction parameter $g$ was estimated independently using a simple model \cite{levy2012luttinger},

\begin{equation}
g = \left(1 + \frac{U}{2E_F}\right)^{-1/2},
\end{equation}

where $U$ is the Coulomb potential and $E_F$ is the Fermi energy. The Coulomb energy is approximated as

\begin{equation}
U = \frac{e^2}{4\pi \varepsilon_0 r},
\end{equation}

with $r \sim 1/n$ in a one-dimensional system. For the first subband, the density is estimated as $n = 3 \times 10^{8}\,\mathrm{m^{-1}}$ \cite{makaju2024nonreciprocal}. The Fermi energy is given by

\begin{equation}
E_F = \frac{\hbar^2 k_F^2}{2m^*}, 
\qquad 
k_F = \frac{\pi n}{2}.
\end{equation}

Using these, eq.3 yields $g = 0.606$, independent of the fitting procedure.

Among the three models, the logarithmic form provides the best statistical fit to 
the data. However, it yields a very low characteristic temperature, suggesting that 
this dependence should be most meaningful only at very low temperatures. For the strong-disorder model \cite{kane1992transmission},

\begin{equation}
G = A \left(\frac{V}{V_0}\right)^{\beta},
\end{equation}

where $\beta = \frac{1}{g} - 1$, $G$ is the conductance, $A$ is a fitting constant, 
and $V_0$ defines an energy scale

\begin{equation}
T_0 = \frac{e V_0}{k_B}.
\end{equation}

For the first subband, we obtain g = 0.916, $\beta$ = 0.09114 and $T_0$ = 8.78 mK. The extracted interaction parameter is close to unity, indicating weak electron–electron interactions in the first subband. The corresponding energy scale $T_0$ suggests moderate disorder strength. However, this fit matches the data less closely, with noticeable discrepancies at both low and high applied biases.

Finally, for the weak-disorder model \cite{maslov1995transport},

\begin{equation}
G = g_0 \left[ 1 - \left(\frac{V}{V_0}\right)^{\alpha} \right],
\end{equation}

where $\alpha = 1 - g$ and $g_0 = 2e^2/h$. From this fit we extract g = -0.92, $\alpha$ = 1.092 and $T_0$ = 2.34 mK. Although this model provides a reasonable description of the bias dependence, the extracted interaction parameter has an unphysical sign, inconsistent with theoretical expectations for repulsive interactions in a one-dimensional electron 
system \cite{levy2006luttinger,maslov1995transport}.

Taken together, these results indicate that no single model provides a fully 
self-consistent microscopic description of the data. The logarithmic form best captures the observed bias dependence. However, the conductance remains near quantized values and the extracted $T_c$ is very low, making it unlikely that the system is in a strongly localized regime. The strong-disorder model yields interaction parameters closer to theoretical 
expectations, but it does not reproduce the data particularly well, and transport 
through a $7\,\mu\mathrm{m}$ wire is unlikely to be described by a single-barrier model. The weak-disorder model, while qualitatively consistent with interaction-driven 
suppression, produces unphysical parameter values. 

This comparison highlights the complex interplay between disorder and 
electron–electron interactions in our system and suggests that a simple 
barrier transmission model is insufficient to fully describe transport in 
these wires. Nevertheless, the extracted characteristic energy scales 
ranging between $1.8$ and $8.8\,\mathrm{mK}$, consistent with increasing disorder effects observed in our wire at $V_{\mathrm{bias}} = 0.92\,\upmu\mathrm{V} = 10.7\,\mathrm{mK}$. 
Fits to higher subbands, presented in Supplementary Section 5, yield similar conclusions. Similar results were also obtained by simulating the transmission through two asymmetric barriers, as shown in Section 6 of the Supplement. These fittings, consistent with large potential barriers forming at the intersection of the 1D and the 2D sections of the device, yielded a largest barrier height of 1.61 $\upmu$V = 18.7 mK and width of 460 nm.  

In summary, we report 1D-Coulomb drag measurements in a device with a ballistic quantum wire. Despite the drag wire having minimal disorder scattering, the non-reciprocal component of the drag signal still has a strength comparable to that of the reciprocal component, which is unexpected under the charge-fluctuation formalism. This discrepancy might be due to the low-energy disorder within the system, and the behavior of the low-energy wire conductance is consistent with this assumption. Both the temperature dependence of the drag signal and the bias voltage dependence of the conductance cannot be fully understood within currently existing theoretical models, highlighting the importance of continuous theoretical and experimental efforts in this field.

\begin{acknowledgments}
This work was supported by the National Science Foundation through NSF/DMR-2518016. This work was performed, in part, at the Center for Integrated Nanotechnologies, an Office of Science User Facility operated for the U.S. Department of Energy (DOE) Office of Science. Sandia National Laboratories is a multimission laboratory managed and operated by National Technology $\And$ Engineering Solutions of Sandia, LLC, a wholly owned subsidiary of Honeywell International, Inc., for the U.S. DOE’s National Nuclear Security Administration under contract DE-NA-0003525. The views expressed in the article do not necessarily represent the views of the U.S. DOE or the United States Government. Part of this work was also conducted at the Research Service Centers (RSC) of the Herbert Wertheim College of Engineering at the University of Florida.
\end{acknowledgments}

\section*{Data availability}
The data that support the findings of this study are available within the paper and
its Supplementary Information. The raw data that support the findings of this study
will be available in Zenodo upon publication.

\section*{Author Contributions} 
S.J.A. grew the GaAs/AlGaAs heterostructure. R.M., M.Z. and D.L. fabricated the device in the RSC. S.C. and R.M. took the measurements. S.C. analyzed the data. N.R. and G.G. performed the transmission modeling. D.L. designed and supervised the experiment. S.C. and D.L. co-wrote the Letter, and all authors discussed the results and commented on the manuscript.


\bibliography{apssamp}

@PREAMBLE{
 "\providecommand{\noopsort}[1]{}" 
 # "\providecommand{\singleletter}[1]{#1}%" 
}

@article{maslov1995transport,
  title = {Transport through dirty Luttinger liquids connected to reservoirs},
  author = {Maslov, Dmitrii L.},
  journal = {Phys. Rev. B},
  volume = {52},
  issue = {20},
  pages = {R14368--R14371},
  numpages = {0},
  year = {1995},
  month = {Nov},
  publisher = {American Physical Society},
  doi = {10.1103/PhysRevB.52.R14368},
  url = {https://link.aps.org/doi/10.1103/PhysRevB.52.R14368}
}

@article{kane1992transmission,
  title = {Transmission through barriers and resonant tunneling in an interacting one-dimensional electron gas},
  author = {Kane, C. L. and Fisher, Matthew P. A.},
  journal = {Phys. Rev. B},
  volume = {46},
  issue = {23},
  pages = {15233--15262},
  numpages = {0},
  year = {1992},
  month = {Dec},
  publisher = {American Physical Society},
  doi = {10.1103/PhysRevB.46.15233},
  url = {https://link.aps.org/doi/10.1103/PhysRevB.46.15233}
}

@article{tang2019long,
author = {Tang, Yuhe and Tylan-Tyler, Anthony and Lee, Hyungwoo and Lee, Jung-Woo and Tomczyk, Michelle and Huang, Mengchen and Eom, Chang-Beom and Irvin, Patrick and Levy, Jeremy},
title = {Long-Range Non-Coulombic Electron–Electron Interactions between LaAlO3/SrTiO3 Nanowires},
journal = {Advanced Materials Interfaces},
volume = {6},
number = {15},
pages = {1900301},
doi = {https://doi.org/10.1002/admi.201900301},
url = {https://advanced.onlinelibrary.wiley.com/doi/abs/10.1002/admi.201900301},
year = {2019}
}

@article{borin2019coulomb,
  title = {Coulomb drag effect induced by the third cumulant of current},
  author = {Borin, Artem and Safi, Ines and Sukhorukov, Eugene},
  journal = {Phys. Rev. B},
  volume = {99},
  issue = {16},
  pages = {165404},
  numpages = {9},
  year = {2019},
  month = {Apr},
  publisher = {American Physical Society},
  doi = {10.1103/PhysRevB.99.165404},
  url = {https://link.aps.org/doi/10.1103/PhysRevB.99.165404}
}

@article{gramila1991mutual,
  title = {Mutual friction between parallel two-dimensional electron systems},
  author = {Gramila, T. J. and Eisenstein, J. P. and MacDonald, A. H. and Pfeiffer, L. N. and West, K. W.},
  journal = {Phys. Rev. Lett.},
  volume = {66},
  issue = {9},
  pages = {1216--1219},
  numpages = {0},
  year = {1991},
  month = {Mar},
  publisher = {American Physical Society},
  doi = {10.1103/PhysRevLett.66.1216},
  url = {https://link.aps.org/doi/10.1103/PhysRevLett.66.1216}
}

@article{kamenev1995coulomb,
  title = {Coulomb drag in normal metals and superconductors: Diagrammatic approach},
  author = {Kamenev, Alex and Oreg, Yuval},
  journal = {Phys. Rev. B},
  volume = {52},
  issue = {10},
  pages = {7516--7527},
  numpages = {0},
  year = {1995},
  month = {Sep},
  publisher = {American Physical Society},
  doi = {10.1103/PhysRevB.52.7516},
  url = {https://link.aps.org/doi/10.1103/PhysRevB.52.7516}
}

@article{bockrath1999luttinger,
  title={Luttinger-liquid behaviour in carbon nanotubes},
  author={Bockrath, Marc and Cobden, David H and Lu, Jia and Rinzler, Andrew G and Smalley, Richard E and Balents, Leon and McEuen, Paul L},
  journal={Nature},
  volume={397},
  number={6720},
  pages={598--601},
  year={1999},
  publisher={Nature Publishing Group UK London},
  doi = {https://doi.org/10.1038/17569}
}

@article{zheng2025tunable,
  title={Tunable reciprocal and nonreciprocal contributions to 1D Coulomb drag},
  author={Zheng, Mingyang and Makaju, Rebika and Gazizulin, Rasul and Addamane, Sadhvikas J and Laroche, Dominique},
  journal={Nature Communications},
  volume={16},
  number={1},
  pages={6963},
  year={2025},
  publisher={Nature Publishing Group UK London},
  doi = {https://doi.org/10.1038/s41467-025-62324-6}
}

@article{zheng2025quasi,
  title = {Quasi-1D Coulomb Drag in the Nonlinear Regime},
  author = {Zheng, Mingyang and Makaju, Rebika and Gazizulin, Rasul and Levchenko, Alex and Addamane, Sadhvikas J. and Laroche, Dominique},
  journal = {Phys. Rev. Lett.},
  volume = {134},
  issue = {23},
  pages = {236301},
  numpages = {8},
  year = {2025},
  month = {Jun},
  publisher = {American Physical Society},
  doi = {10.1103/v3dn-bnrp},
  url = {https://link.aps.org/doi/10.1103/v3dn-bnrp}
}

@article{makaju2024nonreciprocal,
  title = {Nonreciprocal Coulomb drag between quantum wires in the quasi-one-dimensional regime},
  author = {Makaju, R. and Kassar, H. and Daloglu, S. M. and Huynh, A. and Laroche, D. and Levchenko, A. and Addamane, S. J.},
  journal = {Phys. Rev. B},
  volume = {109},
  issue = {8},
  pages = {085101},
  numpages = {8},
  year = {2024},
  month = {Feb},
  publisher = {American Physical Society},
  doi = {10.1103/PhysRevB.109.085101},
  url = {https://link.aps.org/doi/10.1103/PhysRevB.109.085101}
}

@article{laroche20141d,
  title={1D-1D Coulomb drag signature of a Luttinger liquid},
  author={Laroche, Dominique and Gervais, G and Lilly, MP and Reno, JL},
  journal={Science},
  volume={343},
  number={6171},
  pages={631--634},
  year={2014},
  publisher={American Association for the Advancement of Science},
  doi = {10.1103/PhysRevB.109.085101}
}

@article{klesse2000coulomb,
  title = {Coulomb drag between quantum wires},
  author = {Klesse, Rochus and Stern, Ady},
  journal = {Phys. Rev. B},
  volume = {62},
  issue = {24},
  pages = {16912--16925},
  numpages = {0},
  year = {2000},
  month = {Dec},
  publisher = {American Physical Society},
  doi = {10.1103/PhysRevB.62.16912},
  url = {https://link.aps.org/doi/10.1103/PhysRevB.62.16912}
}

@article{fuchs2005coulomb,
  title = {Coulomb drag between quantum wires with different electron densities},
  author = {Fuchs, Thomas and Klesse, Rochus and Stern, Ady},
  journal = {Phys. Rev. B},
  volume = {71},
  issue = {4},
  pages = {045321},
  numpages = {6},
  year = {2005},
  month = {Jan},
  publisher = {American Physical Society},
  doi = {10.1103/PhysRevB.71.045321},
  url = {https://link.aps.org/doi/10.1103/PhysRevB.71.045321}
}

@article{auslaender2005spin,
  title={Spin-charge separation and localization in one dimension},
  author={Auslaender, OM and Steinberg, H and Yacoby, Amnon and Tserkovnyak, Y and Halperin, BI and Baldwin, KW and Pfeiffer, LN and West, KW},
  journal={Science},
  volume={308},
  number={5718},
  pages={88--92},
  year={2005},
  publisher={American Association for the Advancement of Science},
  doi = {https://doi.org/10.1126/science.1107821}
}

@article{chang1996observation,
  title = {Observation of Chiral Luttinger Behavior in Electron Tunneling into Fractional Quantum Hall Edges},
  author = {Chang, A. M. and Pfeiffer, L. N. and West, K. W.},
  journal = {Phys. Rev. Lett.},
  volume = {77},
  issue = {12},
  pages = {2538--2541},
  numpages = {0},
  year = {1996},
  month = {Sep},
  publisher = {American Physical Society},
  doi = {10.1103/PhysRevLett.77.2538},
  url = {https://link.aps.org/doi/10.1103/PhysRevLett.77.2538}
}

@article{levy2006luttinger,
  title = {Luttinger-Liquid Behavior in Weakly Disordered Quantum Wires},
  author = {Levy, E. and Tsukernik, A. and Karpovski, M. and Palevski, A. and Dwir, B. and Pelucchi, E. and Rudra, A. and Kapon, E. and Oreg, Y.},
  journal = {Phys. Rev. Lett.},
  volume = {97},
  issue = {19},
  pages = {196802},
  numpages = {4},
  year = {2006},
  month = {Nov},
  publisher = {American Physical Society},
  doi = {10.1103/PhysRevLett.97.196802},
  url = {https://link.aps.org/doi/10.1103/PhysRevLett.97.196802}
}

@article{microsoft2025interferometric,
  title={Interferometric single-shot parity measurement in InAs--Al hybrid devices},
  author={Microsoft Azure Quantum and Aghaee, Morteza and Alcaraz Ramirez, Alejandro and Alam, Zulfi and Ali, Rizwan and Andrzejczuk, Mariusz and Antipov, Andrey and Astafev, Mikhail and Barzegar, Amin and Bauer, Bela and others},
  journal={Nature},
  volume={638},
  number={8051},
  pages={651--655},
  year={2025},
  publisher={Nature Publishing Group UK London},
  doi = {https://doi.org/10.1038/s41586-024-08445-2}
}

@article{franklin2022carbon,
  title={Carbon nanotube transistors: Making electronics from molecules},
  author={Franklin, Aaron D and Hersam, Mark C and Wong, H-S Philip},
  journal={Science},
  volume={378},
  number={6621},
  pages={726--732},
  year={2022},
  publisher={American Association for the Advancement of Science},
  doi= {https://doi.org/10.1126/science.abp8278}
}

@article{lal2001transport,
  title = {Transport through Quasiballistic Quantum Wires: The Role of Contacts},
  author = {Lal, Siddhartha and Rao, Sumathi and Sen, Diptiman},
  journal = {Phys. Rev. Lett.},
  volume = {87},
  issue = {2},
  pages = {026801},
  numpages = {4},
  year = {2001},
  month = {Jun},
  publisher = {American Physical Society},
  doi = {10.1103/PhysRevLett.87.026801},
  url = {https://link.aps.org/doi/10.1103/PhysRevLett.87.026801}
}

@article{tarucha1995reduction,
  title={Reduction of quantized conductance at low temperatures observed in 2 to 10 $\mu$m-long quantum wires},
  author={Tarucha, Seigo and Honda, Takashi and Saku, Tadashi},
  journal={Solid state communications},
  volume={94},
  number={6},
  pages={413--418},
  year={1995},
  publisher={Elsevier},
  doi = {https://doi.org/10.1016/0038-1098(95)00102-6}
}

@article{levy2012luttinger,
  title = {Experimental evidence for Luttinger liquid behavior in sufficiently long GaAs V-groove quantum wires},
  author = {Levy, E. and Sternfeld, I. and Eshkol, M. and Karpovski, M. and Dwir, B. and Rudra, A. and Kapon, E. and Oreg, Y. and Palevski, A.},
  journal = {Phys. Rev. B},
  volume = {85},
  issue = {4},
  pages = {045315},
  numpages = {5},
  year = {2012},
  month = {Jan},
  publisher = {American Physical Society},
  doi = {10.1103/PhysRevB.85.045315},
  url = {https://link.aps.org/doi/10.1103/PhysRevB.85.045315}
}

@article{steinberg2008charge,
  title={Charge fractionalization in quantum wires},
  author={Steinberg, Hadar and Barak, Gilad and Yacoby, Amir and Pfeiffer, Loren N and West, Ken W and Halperin, Bertrand I and Le Hur, Karyn},
  journal={Nature Physics},
  volume={4},
  number={2},
  pages={116--119},
  year={2008},
  publisher={Nature Publishing Group UK London},
  doi = {https://doi.org/10.1038/nphys810}
}

@article{klaassen2025realization,
  title={Realization of a one-dimensional topological insulator in ultrathin germanene nanoribbons},
  author={Klaassen, Dennis J and Eek, Lumen and Rudenko, Alexander N and van’t Westende, Esra D and Castenmiller, Carolien and Zhang, Zhiguo and de Boeij, Paul L and van Houselt, Arie and Ezawa, Motohiko and Zandvliet, Harold JW and others},
  journal={Nature communications},
  volume={16},
  number={1},
  pages={2059},
  year={2025},
  publisher={Nature Publishing Group UK London},
  doi = {https://doi.org/10.1038/s41467-025-57147-4}
}

@article{yacoby1996nonuniversal,
  title = {Nonuniversal Conductance Quantization in Quantum Wires},
  author = {Yacoby, A. and Stormer, H. L. and Wingreen, Ned S. and Pfeiffer, L. N. and Baldwin, K. W. and West, K. W.},
  journal = {Phys. Rev. Lett.},
  volume = {77},
  issue = {22},
  pages = {4612--4615},
  numpages = {0},
  year = {1996},
  month = {Nov},
  publisher = {American Physical Society},
  doi = {10.1103/PhysRevLett.77.4612},
  url = {https://link.aps.org/doi/10.1103/PhysRevLett.77.4612}
}

@article{levchenko2008coulomb,
  title = {Coulomb Drag at Zero Temperature},
  author = {Levchenko, Alex and Kamenev, Alex},
  journal = {Phys. Rev. Lett.},
  volume = {100},
  issue = {2},
  pages = {026805},
  numpages = {4},
  year = {2008},
  month = {Jan},
  publisher = {American Physical Society},
  doi = {10.1103/PhysRevLett.100.026805},
  url = {https://link.aps.org/doi/10.1103/PhysRevLett.100.026805}
}

@article{luttinger1963exactly,
  title={An exactly soluble model of a many-fermion system},
  author={Luttinger, JM},
  journal={Journal of mathematical physics},
  volume={4},
  number={9},
  pages={1154--1162},
  year={1963},
  publisher={American Institute of Physics},
  doi = {https://doi.org/10.1063/1.1704046}
}

@article{tomonaga1950remarks,
  title={Remarks on Bloch's method of sound waves applied to many-fermion problems},
  author={Tomonaga, Sin-itiro},
  journal={Progress of Theoretical Physics},
  volume={5},
  number={4},
  pages={544--569},
  year={1950},
  publisher={Oxford University Press},
  doi = {https://doi.org/10.1143/ptp/5.4.544}
}

@article{jompol2009probing,
  title={Probing spin-charge separation in a Tomonaga-Luttinger liquid},
  author={Jompol, Yodchay and Ford, CJB and Griffiths, JP and Farrer, I and Jones, GAC and Anderson, D and Ritchie, DA and Silk, TW and Schofield, AJ},
  journal={Science},
  volume={325},
  number={5940},
  pages={597--601},
  year={2009},
  publisher={American Association for the Advancement of Science},
  doi = {https://doi.org/10.1126/science.1171769}
}

@article{onsager1931reciprocal,
  title = {Reciprocal Relations in Irreversible Processes. I.},
  author = {Onsager, Lars},
  journal = {Phys. Rev.},
  volume = {37},
  issue = {4},
  pages = {405--426},
  numpages = {0},
  year = {1931},
  month = {Feb},
  publisher = {American Physical Society},
  doi = {10.1103/PhysRev.37.405},
  url = {https://link.aps.org/doi/10.1103/PhysRev.37.405}
}

@article{sanchez2010mesoscopic,
  title = {Mesoscopic Coulomb Drag, Broken Detailed Balance, and Fluctuation Relations},
  author = {S\'anchez, Rafael and L\'opez, Rosa and S\'anchez, David and B\"uttiker, Markus},
  journal = {Phys. Rev. Lett.},
  volume = {104},
  issue = {7},
  pages = {076801},
  numpages = {4},
  year = {2010},
  month = {Feb},
  publisher = {American Physical Society},
  doi = {10.1103/PhysRevLett.104.076801},
  url = {https://link.aps.org/doi/10.1103/PhysRevLett.104.076801}
}

@article{yamamoto2006negative,
  title={Negative Coulomb drag in a one-dimensional wire},
  author={Yamamoto, M and Stopa, M and Tokura, Y and Hirayama, Y and Tarucha, S},
  journal={Science},
  volume={313},
  number={5784},
  pages={204--207},
  year={2006},
  publisher={American Association for the Advancement of Science},
  doi = {https://doi.org/10.1126/science.1126601}
}

@article{laroche2011positive,
  title={Positive and negative Coulomb drag in vertically integrated one-dimensional quantum wires},
  author={Laroche, D and Gervais, G and Lilly, MP and Reno, JL},
  journal={Nature nanotechnology},
  volume={6},
  number={12},
  pages={793--797},
  year={2011},
  publisher={Nature Publishing Group},
  doi = {https://doi.org/10.1038/nnano.2011.182}
}

@article{tabatabaei2020andreev,
  title = {Andreev-Coulomb Drag in Coupled Quantum Dots},
  author = {Tabatabaei, S. Mojtaba and S\'anchez, David and Yeyati, Alfredo Levy and S\'anchez, Rafael},
  journal = {Phys. Rev. Lett.},
  volume = {125},
  issue = {24},
  pages = {247701},
  numpages = {6},
  year = {2020},
  month = {Dec},
  publisher = {American Physical Society},
  doi = {10.1103/PhysRevLett.125.247701},
  url = {https://link.aps.org/doi/10.1103/PhysRevLett.125.247701}
}

@article{tang2020frictional,
  title={Frictional drag between superconducting LaAlO3/SrTiO3 nanowires},
  author={Tang, Yuhe and Lee, Jung-Woo and Tylan-Tyler, Anthony and Lee, Hyungwoo and Tomczyk, Michelle and Huang, Mengchen and Eom, Chang-Beom and Irvin, Patrick and Levy, Jeremy},
  journal={Semiconductor Science and Technology},
  volume={35},
  number={9},
  pages={09LT01},
  year={2020},
  publisher={IOP Publishing},
  doi = {10.1088/1361-6641/ab9ec9}
}

@article{kaasbjerg2016correlated,
  title = {Correlated Coulomb Drag in Capacitively Coupled Quantum-Dot Structures},
  author = {Kaasbjerg, Kristen and Jauho, Antti-Pekka},
  journal = {Phys. Rev. Lett.},
  volume = {116},
  issue = {19},
  pages = {196801},
  numpages = {6},
  year = {2016},
  month = {May},
  publisher = {American Physical Society},
  doi = {10.1103/PhysRevLett.116.196801},
  url = {https://link.aps.org/doi/10.1103/PhysRevLett.116.196801}
}

@article{narozhny2016coulomb,
  title = {Coulomb drag},
  author = {Narozhny, B. N. and Levchenko, A.},
  journal = {Rev. Mod. Phys.},
  volume = {88},
  issue = {2},
  pages = {025003},
  numpages = {55},
  year = {2016},
  month = {May},
  publisher = {American Physical Society},
  doi = {10.1103/RevModPhys.88.025003},
  url = {https://link.aps.org/doi/10.1103/RevModPhys.88.025003}
}

@article{voit1995one,
  title={One-dimensional Fermi liquids},
  author={Voit, Johannes},
  journal={Reports on Progress in Physics},
  volume={58},
  number={9},
  pages={977--1116},
  year={1995},
  doi = {10.1088/0034-4885/58/9/002}
}

@article{pustilnik2003coulomb,
  title = {Coulomb Drag by Small Momentum Transfer between Quantum Wires},
  author = {Pustilnik, M. and Mishchenko, E. G. and Glazman, L. I. and Andreev, A. V.},
  journal = {Phys. Rev. Lett.},
  volume = {91},
  issue = {12},
  pages = {126805},
  numpages = {4},
  year = {2003},
  month = {Sep},
  publisher = {American Physical Society},
  doi = {10.1103/PhysRevLett.91.126805},
  url = {https://link.aps.org/doi/10.1103/PhysRevLett.91.126805}
}

@article{nazarov1998current,
  title = {Current Drag in Capacitively Coupled Luttinger Constrictions},
  author = {Nazarov, Yuli V. and Averin, D. V.},
  journal = {Phys. Rev. Lett.},
  volume = {81},
  issue = {3},
  pages = {653--656},
  numpages = {0},
  year = {1998},
  month = {Jul},
  publisher = {American Physical Society},
  doi = {10.1103/PhysRevLett.81.653},
  url = {https://link.aps.org/doi/10.1103/PhysRevLett.81.653}
}

@article{anderson2021coulomb,
  title = {Coulomb Drag between a Carbon Nanotube and Monolayer Graphene},
  author = {Anderson, Laurel and Cheng, Austin and Taniguchi, Takashi and Watanabe, Kenji and Kim, Philip},
  journal = {Phys. Rev. Lett.},
  volume = {127},
  issue = {25},
  pages = {257701},
  numpages = {6},
  year = {2021},
  month = {Dec},
  publisher = {American Physical Society},
  doi = {10.1103/PhysRevLett.127.257701},
  url = {https://link.aps.org/doi/10.1103/PhysRevLett.127.257701}
}

@article{xiang2020one,
  title={One-dimensional van der Waals heterostructures},
  author={Xiang, Rong and Inoue, Taiki and Zheng, Yongjia and Kumamoto, Akihito and Qian, Yang and Sato, Yuta and Liu, Ming and Tang, Daiming and Gokhale, Devashish and Guo, Jia and others},
  journal={Science},
  volume={367},
  number={6477},
  pages={537--542},
  year={2020},
  publisher={American Association for the Advancement of Science},
  doi = {https://doi.org/10.1126/science.aaz2570}
}

@article{zheng2026quasi,
  title = {Quasi-one-dimensional Coulomb drag between spin-polarized quantum wires},
  author = {Zheng, Mingyang and Makaju, Rebika and Gazizulin, Rasul and Levchenko, Alex and Addamane, Sadhvikas J. and Laroche, Dominique},
  journal = {Phys. Rev. B},
  volume = {113},
  issue = {12},
  pages = {L121408},
  numpages = {7},
  year = {2026},
  month = {Mar},
  publisher = {American Physical Society},
  doi = {10.1103/5yqq-cw73},
  url = {https://link.aps.org/doi/10.1103/5yqq-cw73}
}

\end{document}